# entropy

*Review*

# Energy Flows in Low-entropy Complex Systems

**Eric J. Chaisson** [1],*

[1] Harvard-Smithsonian Center for Astrophysics, Harvard University, Cambridge, Massachusetts, 02138, USA

* E-Mail: ejchaisson@cfa.harvard.edu
  Tel.: +1-978-505-2667.

Academic Editor:

**Abstract:** Nature's many complex systems—physical, biological, and cultural—are islands of low-entropy order within increasingly disordered seas of surrounding, high-entropy chaos. Energy is a principal facilitator of the rising complexity of all such systems in the expanding Universe, including galaxies, stars, planets, life, society, and machines. A large amount of empirical evidence—relating neither entropy nor information, rather energy—suggests that an underlying simplicity guides the emergence and growth of complexity among many known, highly varied systems in the 14-billion-year-old Universe, from big bang to humankind. Energy flows are as centrally important to life and society as they are to stars and galaxies. In particular, the quantity energy rate density—the rate of energy flow per unit mass—can be used to explicate in a consistent, uniform, and unifying way a huge collection of diverse complex systems observed throughout Nature. Operationally, those systems able to utilize optimal amounts of energy tend to survive and those that cannot are non-randomly eliminated.

**Keywords:** complexity; cosmic evolution; cosmology; energy; energy rate density; evolution; thermodynamics

**PACS Codes :**

## 1. Introduction

This article risks alienating both entropy specialists and information theorists. By contrast, when addressing complexity science generally and practically, I advocate the role of energy, specifically energy flow and its hypothesized complexity metric, "energy rate density." Much research has shown that the non-equilibrium thermodynamics of complex systems can be modeled consistently and uniformly by quantitatively analyzing the way energy is acquired, stored, and expressed by a wide array of systems observed throughout Nature [1, 2, 3].

Here, I briefly review the application of energy rate density to a cosmological selection of complex systems, including galaxies, stars, planets, life, society, and machines. As open, organized, non-equilibrated systems evolve, energy flows can be used to track both the rise of their complexity as well as their evolutionary milestones for an extremely broad spectrum of physical, biological, and cultural systems that have emerged over the course of nearly 14 Gy. With a working definition for complexity in mind—*a state of intricacy, complication, variety or involvement among the networked, interacting parts of a system's structure and function; operationally, a measure of the information needed to describe a system, or of the rate of energy flowing through a system of given mass*—I propose, after much research, discussion, measurement and computation, that energy rate density is potentially a more useful complexity metric (or proxy for it) than either entropy production or information content.

All the complex systems examined below obey the known principles of modern thermodynamics, including its esteemed 2$^{nd}$ law, so readers can be assured that this work accords with fundamental physics. Although every complex system does display lowered entropy, the environments surrounding those systems increase in entropy. And when both individual systems and their environments are considered together, the net change in entropy always rises, in accord with the widely accepted idea that the Universe itself grows increasingly chaotic and equilibrated. Visualizing complex systems as "islands of order" within a "sea of disorder," both order and entropy can increase together—order locally (within systems) and entropy globally (in their surrounding environments).

No new science beyond non-equilibrium thermodynamics is needed to describe *generally* the growth of complexity over the entire course of natural history, from the big bang in the early Universe to humankind on Earth. The novelty discussed herein regards the application of energy rate density in the modeling of complex systems. This article suggests that this quantity can usefully and robustly characterize physical, biological, and cultural systems. Much empirical data for this proposed metric are compactly reviewed below—some measured, others calculated—in order to test the idea that all known complex systems can be modeled in a common, objective, unified way, regardless of whether alive or not.

Energy rate density offers more than an effective metric, capable of assessing structural and functional complexity in thermodynamic systems. It potentially provides a new way to broaden the concept of evolution and to unify the natural sciences, including the physical evolution of relatively simple inanimate systems, the biological evolution of more complex life-forms, and the cultural evolution of some of the most complex systems built by human society. The sum total of these three phases of evolution, broadly conceived, is the interdisciplinary subject of cosmic evolution, a grand scientific narrative now under development (e.g., [3, 4]).

## 2. Cosmic Evolution

An extensive, interdisciplinary synthesis linking numerous specialties—physics, astronomy, geology, chemistry, biology, and anthropology—is now emerging from a thermodynamic analysis of complex systems. This is a coherent, cosmological survey of natural history based on the concept of ubiquitous change—a narrative story that includes our Milky Way, our Sun, our Earth, and ourselves. Often termed cosmic evolution and sometimes big history or the epic of evolution, its definition is broad and inclusive: *Cosmic evolution is the study of the many varied developmental and generational changes in the assembly and composition of radiation, matter, and life throughout the history of the Universe*. Accordingly, an appreciation for evolution, generally considered, potentially extends well beyond the subject of biology, in keeping with the definition of evolution in most dictionaries as *any process of ascent with change in the formation, growth, and development of systems*.

The intent of my research agenda is not to reduce biology to physics. Rather, I seek to broaden physics to include biology—to treat life, like non-life, as a natural consequence of increasingly ordered structures with an expanding, non-equilibrated Universe [5]. Like much else in Nature, complex systems are characterized neither entirely by reductionism nor wholly by holism, any more than solely by chance or necessity. Events at work in the messy, real world—in contrast to abstract, theoretical studies of idealized, textbook systems—resemble syntheses of dichotomies; uncertain randomness mixes with strict determinism, global networking integrates with reductive specialization.

When eclectically examined, evolution crosses all disciplinary boundaries. The most familiar kind of evolution—biological evolution or neo-Darwinism—is just one, albeit important, subset of broader evolutionary action encompassing much more than mere life on Earth. In short, what Darwinian change does for plants and animals, cosmic evolution aspires to do for all material systems. And if Darwinism created a revolution of understanding that humans are no different from other life-forms on our planet, then cosmic evolution extends the simple, yet powerful, idea of change writ large by treating Earth and our bodies in much the same way as stars and galaxies far beyond.

The arrow of time sketched in Figure 1 provides a convenient graphic outlining the sequence of main events that have changed systems throughout natural history from simplicity to complexity, from inorganic to organic, from chaos to order. That sequence, as determined by a huge amount of data collected since Renaissance times, accords well with the idea that a thread of change links, in turn:

- the primal energy that produced elementary particles,
- particles that combined into atoms,
- atoms that clustered into galaxies and stars,
- stars that fused heavy elements,
- elements that synthesized molecular ingredients of life,
- complex molecules that spawned life itself,
- and life that became intelligent within our human society.

Not that time's arrow implies that "lower," primitive life-forms biologically change directly into "higher," advanced organisms, any more than galaxies physically change into stars, or stars into planets. Rather, over the course of historical time the environmental conditions ripened for galactic formation, but now those conditions are more conducive to stellar and planetary formation; likewise more recently, environments suitable for spawning simple life eventually changed into those favoring the emergence of more complex species. The resulting system changes, especially as described by their energy budgets, have *generally* been toward greater amounts of diverse complexity, as numerically condensed in Section 5.

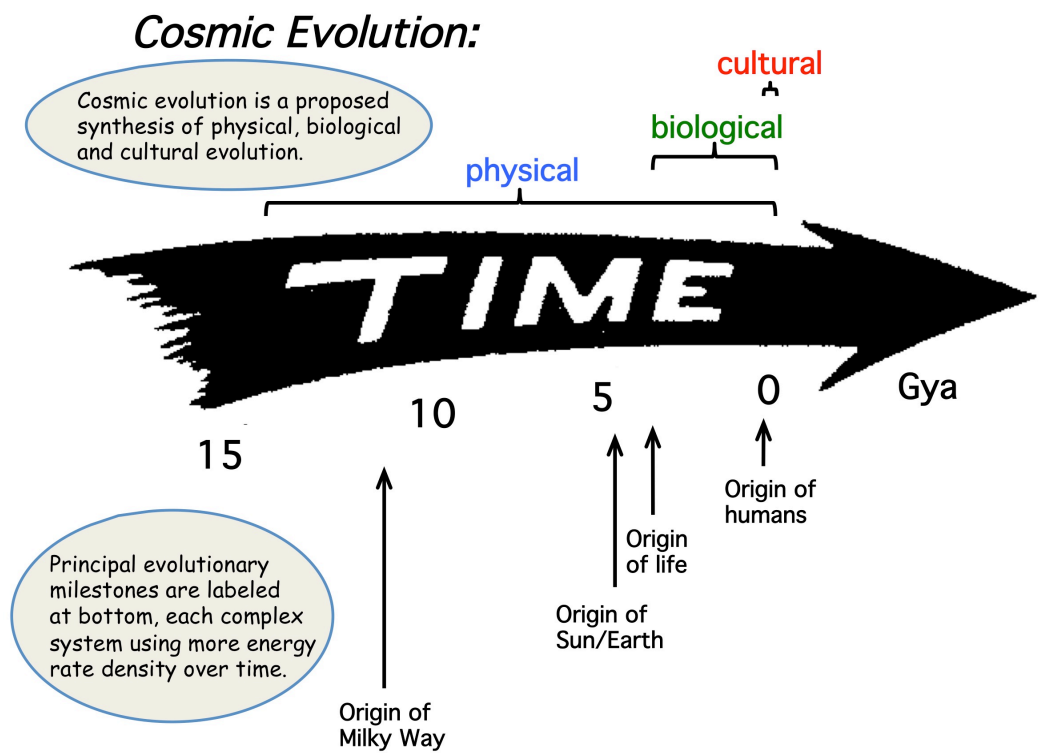

**Figure 1.** An "arrow of time" symbolically charts the grand sweep of cosmic evolution that generally integrates the three phases of physical, biological, and cultural evolution (top), extending over ~14 billion years from the big bang (left) to the present (right). (The curly bracket for cultural evolution at top right would actually be a mere point, if drawn to scale.) Some milestones in the history of our galaxy, star, and planet, as well as of life, humanity, and civilization are labeled (bottom). Despite the arrowhead, no directionality is stated or implied for the evolutionary process, other than that the Universe grows evermore disordered in accord with thermodynamics' 2[nd] law. Nor is there any purpose, plan or design evident in the data supporting the cosmic-evolutionary scenario.

## 3. Entropy and Information

I do not wish to be unnecessarily controversial. I know that many readers of this journal work regularly and well with both entropy and information. I respect and understand that these terms are often useful in probing a wealth of specialized topics. My interests, however, are more general; I seek practical, quantitative analyses applicable to the widest possible spectrum of natural systems.

This is not a criticism of colleagues who study complexity and evolution by employing information theory or entropy production [6, 7]. However, I personally find these methods overly abstract, hard to define (to everyone's satisfaction), and even harder to measure. Regarding entropy, neither maximum nor minimum entropy principles are evident in the data presented in Section 5; nor does the concept of negentropy help. Regarding information, I sense, but cannot prove, that information is another kind of energy; both information storage and retrieval require energy, and greater information processing and calculation need high energy density—but this is a taxing topic for another, separate paper. While information content and entropy estimates offer much theoretical insight, neither provides clear, unambiguous, empirical metrics. I am not alone in this critique; a leading biologist recently advised that both entropy and information be "banned from interdisciplinary discussions of complexity in the history of the Universe" [8].

Entropy does provide useful reality checks that disorder rises within global environments engulfing complex systems per se, thus ensuring that the 2$^{nd}$ law is not violated. This I have done in prior publications (e.g., [1]), where the gain of order within increasingly organized, localized systems can be computed to show how such systems grow and complexify, although only temporarily amidst chaotic surroundings. However, practical ways to empirically test such entropy decreases within complex systems are not as readily available as are measures of energy and energy flows through those systems, as argued in Section 4 and discussed in Section 5.

Nor is information practically useful in quantifying complexity writ large. In biology alone, researchers cannot agree on a complexity metric, despite the recent revolution in genomics and informatics. Some use numerical genome size [9], others gauge body morphology and functional flexibility [10], still others count cell types in organisms [11, 12], chart cellular specialization among species [13], or appeal to networks of ecological interactions [14]. Some of these attributes of life have qualitative worth, notably numbers of different part types [15], yet few hold quantitatively over many levels and scales in biology. To give a single example among morphologically primitive organisms, such as sponges and pre-metazoans, meager cell types often differ dramatically with their genomic wealth [16]. Advanced organisms show similar counter trends; humans' 3.2 billion base pairs well exceed that of a pufferfish (~365 million) yet are greatly exceeded by closely related lungfish (~133 billion); even the wheat genome, which is arguably the most important plant to humans, is at ~17 gigabases several times the size of our human genome. Likewise humans' ~22,300 genes are dwarfed by the ~33,000 genes in a scorpion, ~37,000 in a banana, and ~57,000 in an apple. Protein-coding genes and their base pairs might informatically characterize genomes, but they are faulty markers of species complexity.

The Universe is not likely an information-wired machine obeying a fixed, or even evolving, computer program. The ever-changing cosmos seems more an arena for evolution as a winding, rambling, unpredictable process that includes both chance and necessity in the production and maintenance of complex systems. Such frequent, ubiquitous change over very long periods of historical time seems hardly more (yet nothing less) than the natural way that cultural evolution developed beyond biological evolution, which in turn built upon physical evolution before that (*cf.*, Figure 1). Each of these evolutionary phases comprises an integral part of cosmic evolution's larger purview that also operates naturally, as it always has and likely always will, with the irreversible march of time in the expanding Universe.

As a physicist, I prefer terms that I fully understand without much ambiguity or hidden meaning—quantities that can be defined clearly, measured accurately, computed plainly and tested repeatedly. This is a serious challenge for both entropy and information, whose many contending definitions and methods of measurement have not reached consensus within the scientific community. Information theory, in particular, although a useful concept for communications engineering and its allied fields that send bits and bytes along transmission lines, has not been as helpful when modeling complex systems across an extremely wide spectrum of living and non-living systems. I acknowledge that information may aid in the description of some complex systems, but as stressed below energy is likely needed in the origin and operation of all of them.

My research has therefore retreated from both entropy and information as practical means to diagnose complex systems. I judge those terms empirically deficient as complexity metrics when applied in a uniform manner to a wide array of highly diverse systems observed in Nature. Instead, I embrace energy, energy flows, and in particular a specific measure of energy engaged in structured, functioning systems—namely, energy rate density. Energy is well-defined and well-understood among working scientists; energy can be measured directly and its units are explicit (when treated with care, see next section); not least, many researchers have an intuitive feel for energy. All things considered and as demonstrated below, energy rate density has proven surprisingly robust when surveying a huge array of complex systems extant throughout the Universe.

## 4. Energy and Energy Rate Density

Energy is unequivocally defined in physics and so is energy rate density. *Energy is the ability to do work (or to cause change). Energy rate density is the amount of energy available for work while passing through a system per unit time and per unit mass.* These are clean, clear, unambiguous terms and it is a misconception to assert that energy is poorly defined [17]. It is also unhelpful when some experts use power density to denote several different quantities with several different units, even sometimes confusingly mixing units of distinct metric systems [18]. Some dictionaries use the words energy and power synonymously, which is also wrong.

Energy flows caused by the expanding cosmos do seem to be as central and common to the structure and function of all known complex systems as anything yet discovered in Nature. Those (free) energy flows originated with the loss of symmetry and equilibrium between matter and radiation in the earlier Universe, ~$4 \times 10^5$ y after the big bang [19]. It is the optimized use of such energy flows by complex systems, as quantitatively detailed below, that might act as a motor of cosmic evolution on larger scales, thereby affecting physical, biological, and cultural evolution on smaller scales.

Energy not only plays a role in ordering and maintaining complex systems; it might also determine their origin, evolution, and destiny. Recognized decades ago at least qualitatively in words and mostly in biology [20-22], the need for energy is now embraced as an essential organizing feature not only of biological systems such as plants and animals but also of physical systems such as stars and galaxies (*e.g.*, [23-29]). If fusing stars lacked energy flowing within them, they would implode; if plants did not photosynthesize sunlight, they would wilt and die; if humans stopped eating, we too would perish. Energy's essential role is also widely recognized in cultural systems such as a city's inward flow of food and resources amidst its outward flow of products and wastes; energy is a key ingredient in today's economy, technology, and civilization [30].

Energy, therefore, is a quantity that has much commonality among many complex systems. However, the quantity of choice cannot be energy alone, for a star is obviously more energetic than a flower, a galaxy much more energetic than a single cell. By contrast, any living system is surely more complicated than any inanimate entity. Absolute energies are not as indicative of complexity as relative values, which depend on a system's size, composition, coherence, structure, and function. To characterize complexity objectively—that is, to normalize all such complex systems in precisely the same way—a kind of energy density is judged most useful. Moreover, it is the *rate* at which energy transits complex systems of given mass that seems especially constructive (as has long been realized for ecosystems [20, 31]), thereby delineating energy *flow*. Hence, energy rate density, symbolized by $\Phi_m$, is a useful operational term whose expressed intent and plain units are easily understood.

Consider some numerical examples, being careful to get the units correct. Take first our Sun, which has an energy output, or luminosity, in the CGS metric units commonly used by astronomers of $4 \times 10^{33}$ erg/s; its total mass in the same set of units is $2 \times 10^{33}$ g. Equivalently, in Scientifique Internationale units preferred by this journal, those quantities are $4 \times 10^{26}$ J/s and $2 \times 10^{30}$ kg, respectively. Thus, $\Phi_m = 2$ erg/s/g or $2 \times 10^{-4}$ J/kg. This quantity can also be expressed as a power density of $2 \times 10^{-4}$ W/kg, but experience has shown that the term "power" is not as well grasped, even among scientists, as is the explicit term "energy rate"; thus I prefer to be literally clear and transparent in using energy rate density in order to stress that this quantity is a measure of the extent to which energy flows through a system's bulk mass. The above values are correct with their respective units, but caution is advised that some researches label power density oddly and use peculiar, hybrid units. Numerically, energy rate densities expressed in CGS units normally used by most practitioners in the field and those in MKS units required by this journal differ by a factor of $10^4$, the former being larger than the latter.

Consider another typical system, this one alive—in fact, us. In contrast to the physical system above (in fact, any normal star or galaxy), living systems are widely considered to be more complex— and their values of energy rate density, without exception, show that to be the case. Adult human beings, globally averaged today, normally consume ~2800 kcal/day in the form of food to fuel our metabolism. This energy, gained directly from that stored in other (plant and animal) organisms, is sufficient to maintain our bodily structure as well as to drive our physiological functions. Metabolism is a dissipative process—a genuinely thermodynamic mechanism. Heat is generated continuously owing to work done by the tissues among the internal organs of our bodies, including contracting muscles that run the heart, diaphragm, and limbs, ion pumps that maintain the electrical properties of nerves, and biochemical reactions that dismantle food and synthesize new tissue. Therefore, with an average body mass of 65 kg, a generic adult (male or female) maintains $\Phi_m \approx 2$ J/s/kg while in good health. Our metabolic rates, hence values of $\Phi_m$, increase by up to an order of magnitude when performing occupational tasks or recreational activities such as running, jumping, or even writing journal articles; this is how function adds to structure, both requiring energy to sustain them [3]. This is also how humans, like most species of animals, contribute to the rise of entropy in the Universe. We consume high-quality energy in the form of ordered foodstuffs, then radiate away as body heat an equivalent amount of energy as low-quality, disorganized infrared photons. Like stars and galaxies, humans are indeed dissipative structures as are all Earthly life-forms.

Finally, consider a few examples of cultural systems, which *a priori* might be expected to be even more complex than a single life-form. A notable social system such as our human society is, after all, a collection of many individual human beings; and if the whole is truly greater than the sum, society's energy rate density would have among the highest of energy rate densities among all known complex systems—and it does. As modern civilization goes about its daily business, its ~7.3 billion inhabitants utilize ~19 TW to keep its open, ordered, complex society operating [32]; thus $\Phi_m \approx 50$ J/s/kg. Complexity measures for sophisticated, functioning machines, such as automobiles, computers, and aircraft, can reach even an order of magnitude higher, as noted below. Energy rate densities for a large collection of built, cultural systems have been compiled in a study that compares and contrasts the advance of humankind with the rise of machines [33].

Before getting more technical in the next section, note a revealing factoid that many colleagues find surprising. Although the absolute energy of astronomical systems greatly exceeds that of our human selves, and although the mass densities of stars, planets, bodies, and brains are all comparable, the energy rate densities for human beings and our modern society are roughly a million times greater than for stars and galaxies. That's because the quantity $\Phi_m$ is an energy rate *density*. For example, although the Sun emits energy at a prodigious rate, it also contains an unworldly large mass; as just computed above, each second an amount of energy equaling only $2 \times 10^{-4}$ Joule passes through each kilogram of this star. In contrast to any star, more energy (typically a Joule) flows through each kilogram of a plant's leaf during one second of photosynthesis, and much more energy (~20 Joules) pervades each kilogram of gray matter in our brains while thinking for a second.

## 5. Complex Systems

This section applies the term energy rate density, $\Phi_m$, to a wide array of complex systems observed to date in the known Universe. Values of $\Phi_m$ are given for many samples of physical, biological, and cultural systems—and in particular for those systems that specifically led to us, namely the Milky Way Galaxy, Sun, and Earth, as well as life-forms, human society, and our invented machines. These findings are also summarized below in an elaborate graph (Figure 1), which plots consistently and uniformly on a single page the increase of energy rate density with the rise of complexity among a vast collection of non-equilibrated systems.

For this briefest of compact summaries, a half-dozen bullets will suffice to compile relevant numerical values of $\Phi_m$ for numerous complex systems and especially to discern the *general* trend of rising complexity with time. Also given in parentheses are some approximate evolutionary times at which various systems emerged in natural history. A full accounting of the data cited is collected in a recent comprehensive review [3]:

- For physical systems, galaxies have $\Phi_m$ values ($10^{-6}$-$10^{-3}$ J/s/kg) among the lowest of known organized structures (although their dark-matter-dominated dwarf companions range even ten times less, whereas sporadically active galaxies have some ten times more). Galaxies display temporal trends in rising values while developing, such as when our Milky Way increased from $10^{-7}$ to $10^{-3}$ J/s/kg while clustering hierarchically from primordial blobs (~12 Gya) to widespread dwarf galaxies to mature normal status (~10 Gya) to its current state today. By the quantitative complexity metric of energy rate density, galaxies are deemed, despite their oft-claimed majestic splendor, not so overly complex compared to many other forms of organized matter—indeed unequivocally simpler than elaborately structured and exquisitely functioning life-forms.
- Stars, too, adjust their internal states while developing or evolving during one or more generations, their values of $\Phi_m$ rising while complexifying with time. Stellar interiors undergo cycles of nuclear fusion that foster steeper thermal and chemical gradients, resulting in increasingly ordered, differentiated layers of heavy elements within highly evolved stars. Stellar size, color, brightness, and composition all change while slowly altering the structure of every star. For the Sun, specifically, those values increase from about $10^{-4}$ to $10^{-2}$ J/s/kg while changing from a young protostar (~5 Gya) to an aged red giant in the far future, eventually achieving black dwarf status (0 J/s/kg) as its nuclear fires cease, its bulk dissipates, its core shrivels and cools, and its formerly stellar being fades to equilibrated blackness—but not for a time longer than the current age of the Universe.

Although of lesser complexity and longer duration, galaxies are nearly as metabolic and adaptive as any life-form—transacting energy while forming new stars, cannibalizing dwarf galaxies falling in, and dissolving older components within—all the while adjusting their modest organization for greater preservation in response to changing galactic environments. Stars, too, have much in common with life—at least as regards energy flow, material resources, and structural integrity while experiencing change, adaptation, and selection. This is not to say that stars and galaxies are alive, nor that they evolve in the strict and limited biological sense. Most researchers would agree that stars and galaxies at least develop and perhaps evolve over multiple generations as well—as evidenced by systematically rising energy rate density values abridged in the two bullets above.

- In turn, among biological systems on Earth, plants and animals regularly exhibit intermediate values of $\Phi_m$ between 0.1 and 10 J/s/kg. Life is known to operate optimally within certain limits of temperature, pressure, salinity, *etc.*, and so it's not surprisingly that life-forms also have optimal ranges of normalized energy flow above and below which they cannot survive. Plants have values well higher than those for normal stars and typical galaxies, as perhaps best confirmed by the most dominant process in Earth's biosphere—photosynthesis. During the few hundred million years of post-Cambrian times for which data are available, microscopic protists (originating ~470 Mya), followed by gymnosperms (~350 Mya), then angiosperms (~125 Mya), and eventually highly efficient $C_4$ plants (~30 Mya) show clear increases in $\Phi_m$ by about an order of magnitude to ~1 J/s/kg.

- Onward across the bush of life (or the arrow of time)—cells, tissues, organs, organisms—much the same temporal trend of rising energy rate density holds for respiring animals while they evolved and complexified. Values of $\Phi_m$ rose from 0.5 to 10 J/s/kg as adult animals evolved from fish and amphibians (370-500 Mya) to cold-blooded reptiles (~320 Mya) and then to warm-blooded mammals (~200 Mya) and birds in flight (~125 Mya). To the dismay of some observers who feel that humans are special, we are actually typical of mammals and not above them. Not surprisingly, brains have the highest energy rate densities for all living individuals, typically an order of magnitude larger than for the bodies that house them for each and every species examined.

Without exception, system functionality and genetic inheritance—two factors above and beyond mere system structure—help to enhance complexity among animate systems that are clearly living compared to inanimate systems that are clearly not. Energy conceivably acts as a fuel for change, partly and optimally selecting systems able to utilize increased energy rate densities, while forcing others to destruction and extinction—all likely in accord with the widely accepted neo-Darwinian principles of biological evolution. The rise of $\Phi_m$ among plants and animals *generally* parallels the emergence of major evolutionary stages on the scale of life's history. However, given the inexact numbers in the two bullets above, the role of energy rate density in the evolutionary advances along individual biological lineages remains broad and general, not specific and detailed.

- Among cultural systems, advances in technology are comparable to those of human society, each of them energy-rich. Many have values of $\Phi_m$ greater than 100 J/s/kg, hence plausibly the most complex systems known. Social progress can be tracked, again in terms of normalized energy consumption spanning 4 to 200 J/s/kg for a variety of human-related cultural advances among our ancestral forebears—from hunter-gatherers (300,000 ya) to early agriculturists (11,000 ya) to pioneering industrialists (2 centuries ago) to technologists of today.

- Machines, too, and not just computers, but also ordinary engines and smart gadgets integral to our modern, fast-paced economy, can be cast in evolutionary terms—though here the mechanism of change is less Darwinian than Lamarckian, which stresses accumulation of acquired traits. Either way, energy remains an underlying driver of change, and with rapidly accelerating pace displays a steep upward trend in energy rate density ranging 10 to 3000 J/s/kg—from primitive devices of the industrial revolution (150 ya) to pioneering automobiles (a century ago) to impressive airplanes and computers (past half-century) to today's "smart" jet aircraft. All the numbers and computations in this bullet and the one above come from [3].

Human society and its invented machines are among the most energy-laden systems known, hence plausibly the most complex in the Universe explored thus far. Much as for stars, galaxies, and life itself, the winding road to our technological civilization was doubtlessly paved with increased energy density used, or for intelligent life per-capita energy expended. Culturally increasing values of $\Phi_m$ cited in the last two bullets above—whether slow and ancestral such as for controlled fire and tilled land in early agricultural periods, or fast and modern as for powered engines and programmed computers in today's global economy—relate to evolutionary events in which energy flow and cultural selection played significant roles. The cultural evolution of sophisticated technical gadgets vital to our modern economy can be traced by means of their rising $\Phi_m$, as can the entire global economic system across the planet [30].

Threaded within these six bullets, the underlying unifiers so amply evident throughout the big-bang-to-humankind story—evolution and complexity—seem to parallel each another, much as they have all along the arrow of time. Furthermore, all three quantities comprising the essential ideas of this article—evolution, complexity, and energy rate density—seem to be in synchrony at reasonable levels of empirical, quantitative testing.

For those readers preferring words devoid of numbers, a simple "translation" of the bullets above reveals a hierarchically ranked order of increasingly complex systems across all of historical time:

- mature galaxies are more complex than their dwarf predecessors
- red-giant stars are more complex than their main-sequence counterparts
- eukaryotes are more complex than prokaryotes
- plants are more complex than protists
- animals are more complex than plants
- mammals are more complex than reptiles
- brains are more complex than bodies
- societies are more complex than individual humans.

Figure 2 consolidates much recent research on this subject into a single graphic, depicting the transformation of homogeneous, primordial matter (at lower left) into increasingly intricate systems (upper right). Plotted energy rate densities and historical dates are estimates for the general category to which each system belongs. Note that Figures 1 and 2 are drawn with exactly the same horizontal time scale to permit easy comparison; the colors of the insert bubbles in Figure 2 also match those of the evolutionary phases in Figure 1. To display a single, common factor characterizing samples of all known complex systems from big bang to humankind, literally on the same page, is rare even in interdisciplinary science; it should not be overlooked. Energy rate density does seem to be a reasonable candidate to aid unification of the sciences, linking physical, biological, and cultural evolution over ~14 Gy. Such a plot also suggests that an underlying principle, general law, or ongoing process might well be at work—creating, organizing, and maintaining complex systems everywhere.

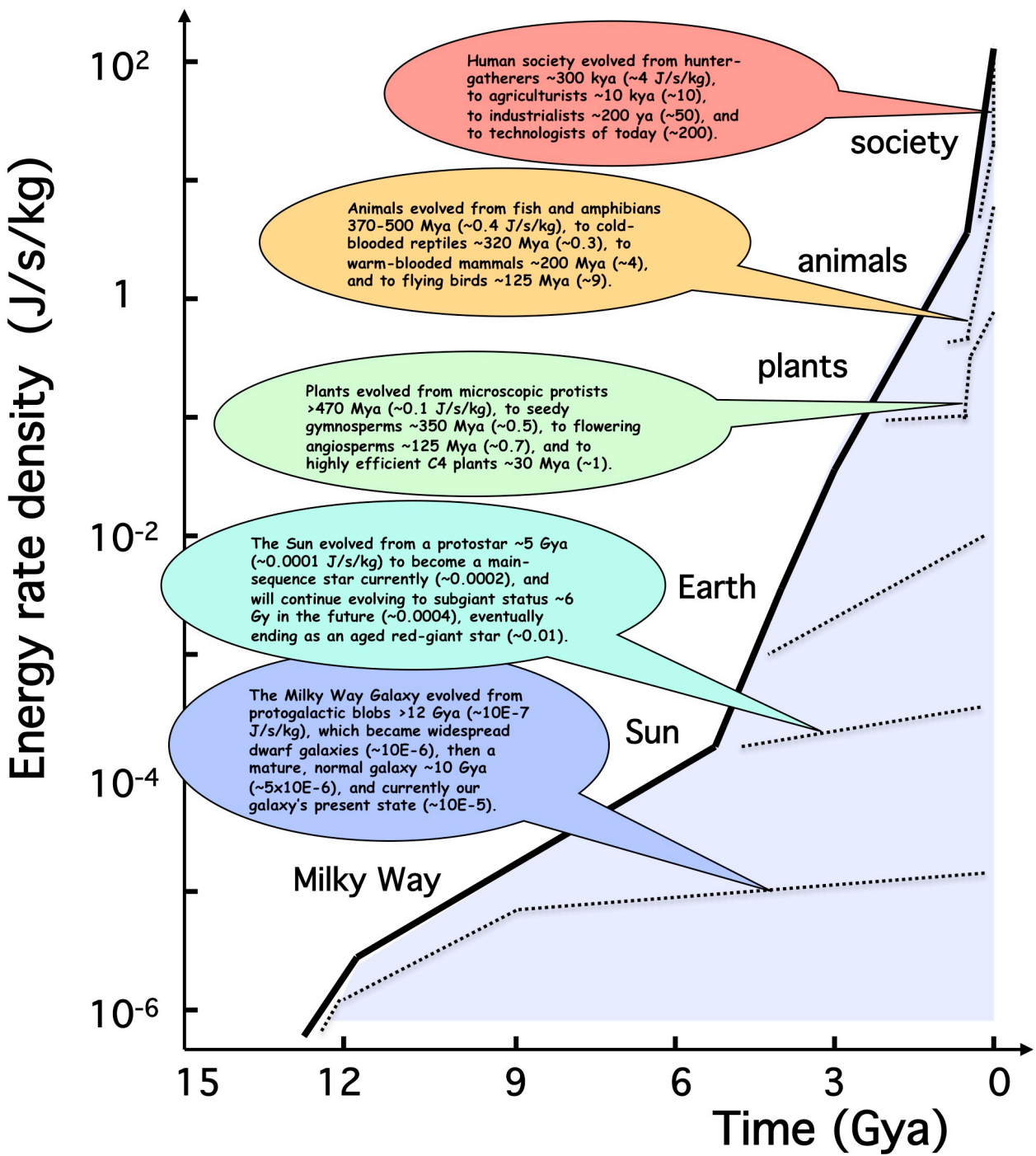

**Figure 2.** Energy rate densities for a wide array of ordered systems that emerged at various times in natural history display a clear trend across ~14 Gy, as simple primordial matter changed into increasingly intricate, complex systems, extending from time's origin (left) to the present (right). The bold black curve of $\Phi_m$ rises roughly exponentially as cultural evolution (steepest slope at upper right) acts faster than biological evolution (moderate slope in middle part of curve), which in turn surpasses physical evolution (smallest slope at lower left). Multiply the vertical axis' values of $\Phi_m$ by $10^4$ to convert SI metric units of J/s/kg to the more commonly used CGS metric units of erg/s/g. The shaded area includes a huge ensemble of evolutionary paths as many varied types of complex systems changed

and complexified since their origin; the several smaller dotted lines sketch some notable evolutionary paths from big bang to humankind, namely, our galaxy, star, plants, animals and society, as compiled in the bubble inserts. The timescale along the abscissa and the color of the bubbles match equivalent parts of Figure 1. Rationale for this plot can be found in an earlier book [1]; updated energies and dates used to make it are taken from a recent review paper [3].

The evidence-based Figure 2 not only encapsulates the physical, biological, and cultural evolution of simple, unorganized matter of the early Universe into ordered systems of growing complexity, but also shows how evolution has done so with accelerating speed, hence the rapidly rising curve. That main, upward trend is represented by the bold black curve extending diagonally across the whole graph. The several smaller dotted lines within the shaded area beneath it sketch some notable evolutionary paths that led to humankind—namely, the Milky Way, Sun, and Earth, as well as much life on our planet, as compiled in the six long bullets above. Many complex systems often show their energy rate density rising exponentially only for limited periods of time, after which their sharp rise tapers off [34-36]. Some slow their rate of growth while following a classic S-shaped curve—much as microbes in a petri dish while replicating unsustainably or as human population is expected to plateau later this century. The bold curve stretching across all of evolutionary history in this figure is probably a compound sum of countless S-curves representing the rise and fall of myriad complex systems over the course of cosmic time [37].

As in any simple, unifying précis of an imperfect Universe—especially one like cosmic evolution that aspires to address all of Nature—there are variations. Exceptions, outliers, or whatever one wants to call those data points that inevitably deviate from the norm, are occasionally evident. However, variation is an essential feature of evolution, on all scales and all times. If there were no variations, adaptation, adjustment, and selection would not work. It is likely that from those variations arose the great diversity among complex, evolving systems—and without them novelty and creativity in the Universe might be absent. Given inevitable uncertainties and variations causing some data to deviate from the mean, bold curve of Figure 2, precise values of the many plotted quantities are not as informative as much as the overall upward trend with the march of time that is so clearly evident in this single graph.

What seems inherently attractive in this analysis is that energy flows caused by the expanding Universe are a central leitmotif regarding the structure and function of all known complex systems. Perhaps Nature's most common currency, energy flows help suppress entropy within increasingly ordered, localized systems evolving amidst increasingly disordered, global environments—a universal process that has arguably governed the emergence and maturity of our galaxy, our star, our planet, and ourselves. Optimality is likely favored in the use of such energy—not so little as to starve a system, yet not so much as to destroy it. Whether stars, life, or civilization, optimal ranges of $\Phi_m$ prevail; too little or too much and systems abort. If correct, energy itself is a key mechanism of change—a central feature of evolution writ large, perhaps even the motor of evolution broadly conceived. And energy rate density is an unambiguous, objective measure of energy flow, enabling us to assess all complex systems in like manner, as well as to gauge how over the course of all cosmic time some systems evolved to command energy and survive, while others apparently could not and did not.

## Acknowledgments

I thank faculty and students at Harvard College Observatory and Smithsonian Astrophysical Observatory who have challenged me with insightful, helpful discussions of the interdisciplinary topic of cosmic evolution. I also appreciate the detailed comments received from two anonymous referees. Support for this informal research program was received in varying degrees over many years from the Sloan Foundation, Smithsonian Institution, Harvard University, National Aeronautics and Space Administration, National Science Foundation, and la Fondation Wright de Geneve.

## Conflicts of Interest

The author declares no conflict of interest.